\documentstyle[preprint,aps]{revtex}
\tightenlines

\begin{document}
\draft

\title{Tuning the spin Hamiltonian of NENP by external pressure: 
a neutron scattering study}

\author{ I.~A.~Zaliznyak$^{1,2}$\cite{Kapitza}, D. C. Dender$^1$\cite{NIST}, 
C. Broholm$^{1,2}$, and Daniel H. Reich$^1$ }
\address{
$^1$Department of Physics and Astronomy, The Johns Hopkins
University, Baltimore, Maryland 21218\\
$^2$National Institute of Standards and Technology, Gaithersburg, Maryland 
20899}

\date{\today}
\maketitle
\widetext
\begin{abstract}

We report an inelastic neutron scattering study of antiferromagnetic spin
dynamics in the Haldane chain compound Ni(C$_2$H$_8$N$_2$)$_2$NO$_2$ClO$_4$
(NENP) under external hydrostatic pressure $P = $ 2.5~GPa. At ambient 
pressure, the magnetic excitations in NENP are dominated by a long-lived triplet  
mode with a gap which is split by  orthorhombic crystalline
anisotropy into a lower doublet centered at $\Delta_\perp\approx$
1.2meV and a singlet at $\Delta_\|\approx$ 2.5meV.  With pressure
we observe appreciable shifts in these levels, which move to
$\Delta_\perp{\rm( 2.5GPa)}\approx$ 1.45~meV and $\Delta_\|{\rm( 2.5GPa)}\approx$
2.2meV. The dispersion of these modes in the crystalline $\hat{c}$-direction
perpendicular to the chain  was measured here for the first time, and can be
accounted for by an  interchain exchange $J'_c \approx 3\times 10^{-4} J$
which changes only slightly with pressure. Since the average gap value 
$\Delta_H\approx$ 1.64~meV remains almost unchanged with $P$, we conclude 
that in NENP the application of external pressure does not affect the 
intrachain coupling $J$ appreciably, but does produce a significant decrease 
of the single-ion anisotropy constant from $D/J = 0.16(2)$ at ambient pressure 
to $D/J = 0.09(7)$ at $P = 2.5 $~GPa.

\vspace{0.5in}


\end{abstract}
%\newpage

\section{INTRODUCTION}

The Haldane ground state \cite{Haldane} with short-range correlations and
a gap to magnetic excitations has  now been found in several 
quasi-one-dimensional S=1 antiferromagnets 
\cite{NENP1,NENP2,YBaNO,AgVP,TMNIN}. 
The magnetic properties of these materials are well-described by
the spin Hamiltonian 

\begin{equation}
\label{H1}
{\cal H} = \sum_{i} \left\{ J {\bf S}_{i} \cdot {\bf S}_{i+1}
+ J' \sum_{\delta_\perp} {\bf S}_{i} \cdot
{\bf S}_{i+\delta_{\perp}} + D(S_{i}^{z})^2 \right\}\;\; ,
\end{equation}
where in addition to  the intrachain and interchain couplings $J$ and $J'$
we have included a uniaxial single-site anisotropy, a common feature of these 
systems. Because of its finite spin-spin correlation length, the Haldane state 
is stable with respect to small perturbations by the interchain exchange and 
single-ion anisotropy, and hence can be found not only in the pure 1D 
Heisenberg limit of Eq.~(\ref{H1}), $J' = D = 0$, but for a finite range of 
$J'$ and $D$ around this point. If, however,  the interchain coupling $J'$ 
exceeds some critical value $J'_{crit.}$, quantum disorder is no longer 
favored and the system enters a long-range ordered N\'{e}el ground state 
with soft Goldstone excitations. Thus, a zero-temperature phase transition 
is expected to occur in a quasi-one-dimensional S=1 antiferromagnet 
at $J' = J'_{crit.}$ if the interchain coupling could be varied. Such quantum
phase transitions have recently attracted much theoretical attention 
\cite{Affleck,Azzouz,Senechal,Botet,SchulzZiman,Sakai,Meshkov,Jolicoeur}. 
Various estimates using different model approximations 
\cite{Affleck,Azzouz,Senechal} as well as numerical studies on finite 
systems \cite{Sakai} have been performed, giving values 
for the critical interchain coupling in the broad range $1.3\times10^{-3}\le 
J'_{crit}/J \le 2.6\times10^{-2}$. On the other hand, solid upper and lower 
limits on $J'_{crit} /J$ are available from  experiments.  It is found that 
the hexagonal antiferromagnet CsNiCl$_3$ with $J\approx$ 2.8~meV and 
$J' /J \approx 2 \times 10 ^{-2}$ undergoes 3D ordering at $T_{\rm N}\approx 
4.8$~K \cite{Yelon-Cox}, while orthorhombic Ni(C$_2$H$_8$N$_2$)$_2$NO$_2$ClO$_4$ 
(NENP) with $J \approx 4$~meV and $J' / J \approx 8 \times 10 ^{-4}$ 
demonstrates all features of the Haldane system and remains disordered down to 
millikelvin temperatures \cite{Meisel}.

The effects of single-ion anisotropy are less drastic than those
of interchain coupling, but are also  eventually critical.
As the presence of the single-ion term in Eq.~(\ref{H1}) does not change 
the basic 1D nature of the model, its effects may be readily explored by 
numerical simulations on  finite systems of realistic size, and have been 
addressed in a number of studies \cite{Botet,SchulzZiman,Sakai,Meshkov,Jolicoeur}. 
These predict that strong Ising-like anisotropy ($D < 0$)  favors N\'{e}el 
ordering for $D < -0.25 J$, while  ``easy-plane'' anisotropy ($D>0$) eventually 
drives the system  into a so-called ``planar'' phase at $D \approx J$ where all 
spins have $S^z =0$ in the ground state.

The currently available experimental data on the effect of nonzero $J'$ and
$D$ on the Haldane spin chain have been obtained from the comparison of
results from different S=1 quasi-1D antiferromagnetic compounds. This restricts
the experimentally accessible region of the ($J',D$) phase diagram discussed
above to just a few points. However, the exchange and anisotropy parameters
of the spin Hamiltonian can be controlled not only by changing the chemical
composition of the magnetic compound, but also by tuning the properties of a 
single material. The application of hydrostatic pressure can change the 
interatomic separations and local atomic environments to which  the spin 
Hamiltonian is highly sensitive, and therefore provides an opportunity 
for the controlled and continuous tuning of the parameters in the Hamiltonian, 
albeit in a restricted range. Here we report a study of the pressure dependence 
of the magnetic properties of NENP. We have used inelastic neutron scattering to
measure the changes in the Haldane gap modes  at $T\approx$1.8 K when an 
external pressure  $P \approx$ 2.5 GPa is applied. We find that both the 
intrachain and interchain coupling are only marginally affected, but
the single-ion anistropy is reduced substantially, and thus hydrostatic
pressure drives this material closer to the ideal Heisenberg model and further 
away from any potential quantum phase transitions.  

\section{PROPERTIES OF NENP AT AMBIENT PRESSURE}

NENP is orthorhombic with space group $Pnma$. It has Ni$^{2+}$ chains stretching 
along the {\bf b} direction with two chains per unit cell staggered in the {\bf a}
direction \cite{Meyer}. 
The ambient-pressure ($P \approx 0.1$~MPa) lattice parameters measured 
in our experiment at $T=1.8$ K after the sample had been pressurized are
$a=15.28$ \AA, $b=10.23$ \AA, and $c=8.096$ \AA. In describing neutron 
scattering experiments, we  refer to wavevector transfer in the corresponding 
reciprocal lattice ${\bf Q} = h{\bf a^{*}} + k{\bf b^{*}} + l{\bf c^{*}} 
\equiv (h, k, l)$. Since the Ni$^{2+}$ ions are displaced by ${\bf b}/2$ along 
the chain, it  is  convenient to refer to the component of wavevector 
transfer along the chain as $\tilde{q} = {\bf Q} \cdot ({\bf b}/2) = k \pi$ 
when discussing the 1D behavior of NENP. At ambient pressure the magnetic 
properties of NENP are well described by the Hamiltonian of Eq.~(\ref{H1}) 
with $J=4.0(2)$ meV,   $D/J=0.16(2)$, and an interchain coupling 
in the $\hat{a}$-direction $J'_a/J=8 \times 10 ^{-4}$. The symmetry axis 
of the single-site anisotropy ($z$--axis) is parallel to the chain axis.
The spin dynamics of NENP under these conditions have been carefully studied 
in a number of neutron-scattering experiments \cite{NENP1,NENP2}. It was shown 
that the principal contribution to the spectral density of spin fluctuations 
${\cal S}({\bf Q},\omega)$ comes from a triplet of long-lived  excitations 
above a gap that follow a dispersion relation with fundamental periodicity 
in $\tilde{q}$ of 2$\pi$. Most of the spectral weight is concentrated in the 
close vicinity of the antiferromagnetic point $|\tilde{q}-\pi |< 2\pi/\xi 
\sim 0.3\pi$ (Here $\xi\sim $6--8 spins is the correlation length in 
the Haldane state). Over the full range $0.3\pi\leq\tilde{q} \leq\pi$
where they have been observed \cite{NENP2}, these excitations are well-described 
by the single mode approximation (SMA):

\begin{equation}
\label{SMA}
{\cal S}^{\alpha\alpha}(\tilde{q},\omega) = -\frac{2}{3}\left( \frac{\langle{\cal H}\rangle
}{L}\right) \frac{1-\cos{\tilde{q}}}{\hbar\omega_\alpha(\tilde{q})}
\delta\left[\hbar\omega - \hbar\omega_\alpha (\tilde{q})\right]\;\; ,
\end{equation}
where $\langle{\cal H}\rangle / L$ is the ground state energy per spin.
The simplest dispersion relation that has the correct periodicity and adequately
fits all the experimental data was found to be \cite{NENP2}

\begin{equation}
\label{hw1}
\hbar\omega_\alpha (\tilde{q}) = \sqrt{\Delta_\alpha^2 + v^2\sin^2
 \tilde{q}+A\sin^2\frac{\tilde{q}}{2}}\;\; 
\end{equation}
with $v = 9.7$~meV, and $A = 34$~meV$^2$.
Instead of a degenerate triplet excitation with a single gap
$\Delta\approx 0.41J\approx 1.64$~meV at $\tilde{q}=\pi$ as expected in the
isotropic case \cite{Meshkov,Jolicoeur}, the planar anisotropy in Eq.~(\ref{H1})
splits the triplet into two branches  with $\Delta_\|\approx $ 2.5~meV and 
$\Delta_\perp\approx $1.2~meV for fluctuations polarized parallel and 
perpendicular to the chain axis, respectively. The lower mode is further split 
by a small orthorhombic anisotropy $\tilde{E}[(S_i^x)^2-(S_i^y)^2]$ with
$\tilde{E} \approx 0.01 J~=~0.04$~meV giving $\Delta_x\approx$ 1.34~meV and
$\Delta_y\approx$ 1.16 meV \cite{NENP1}. This latter splitting is smaller than 
the instrumental resolution of our current experiment, and its effects will 
therefore be neglected in our subsequent discussion. The lower doublet also
shows measurable dispersion along the {\bf a}$^*$ direction, with an effective
bandwidth parameter $\Delta E\approx$ 0.65 meV \cite{NENP1}. This is the
basis of the above estimate of the interchain exchange $J'_a$. The dispersion 
along {\bf c}$^*$ has not been measured prior to this work. 

\section{EXPERIMENTAL PROCEDURE}
Our sample was a 99$\%$ deuterated NENP single crystal of mass 0.27~g, which
was a part of the composite sample used in Ref.~\cite{NENP2}. The experiments
were performed on the SPINS cold neutron triple-axis spectrometer at the 
National Institute of Standards and Technology. The beam divergences 
employed  were 50$^{\prime}/k_i$(\AA$^{-1})-80^{\prime}$ around the 
vertically focussing pyrolytic graphite PG(002) monochromator, and 
$80^{\prime}-240^{\prime}$ around the  PG(002) analyzer. A liquid nitrogen 
cooled Be filter was placed in the scattered beam path. For constant-Q scans 
we fixed the final neutron energy $E_f$ at 5.1~meV. The resulting full width 
at half maximum (FWHM) energy  resolution for incoherent elastic scattering 
was $\Delta \hbar \omega =$ 0.27(1)~meV. Scans with  constant energy transfer 
$\hbar\omega$ = 2~meV were performed with $E_f = $ 4.24~meV to optimize the 
pressure cell transmission for both incident and scattered neutrons.

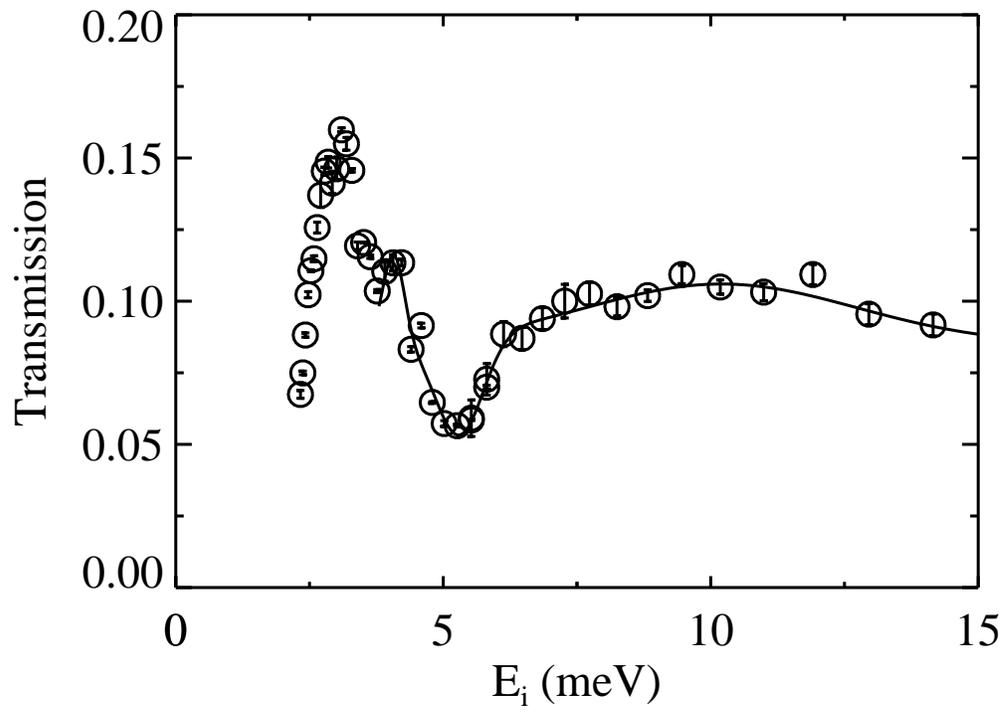
\begin{figure}
\caption{ Energy dependence of neutron transmission for the pressure cell 
loaded with our NENP sample at $T$ = 1.8~K. The solid line is an approximate 
fit given in Eq.~(\protect\ref{transmission}). }
\label{cellTrans}
\end{figure}

To provide high pressures, we used a clamp cell for neutron scattering 
supplied by Oval Co., Ltd., which is described in detail in Ref.~\cite{cell}.
The pressure was generated at room temperature by applying an external load
of 2$\times$10$^5$~N on the cap of the cylindrical sample chamber, or 
``microcell." The microcell is contained inside a barrel-shaped cylinder made 
of high-density polycrystalline Al$_2$O$_3$, which in turn is supported by
steel rings. The microcell accommodated the parallelepiped-shaped
NENP sample (dimensions $\approx$ 6$\times$6$\times$8 mm$^3$), as well
as a  small NaCl platelet (dimensions $\approx 4\times 4\times 2$mm$^3$)
used for pressure calibration. The microcell was filled with the 
pressure-transmitting fluid fluorinert FC-75 (available from 3M Chemicals, Inc.). 
The neutron beam reaches the sample through a 10 mm-high, continuous central 
window between the steel rings after passing through the Al$_2$O$_3$ 
pressure cylinder and the 5 mm-thick Al outer sleeve of the cell. The loaded 
cell was attached to the cold finger of a pumped He$^4$ cryostat for neutron 
scattering and cooled down to 1.8~K. 
Fig.~\ref{cellTrans} shows the transmission through the center of the loaded 
cell at $T=1.8$ K for neutrons in the energy range $2 \le E \le 14$ meV 
(a reduced beam size $\approx 3\times 3$~mm$^2$ was used for this 
measurement). The complicated structure of the curve indicates that the 
transmission is limited by Bragg diffraction from the Al$_2$O$_3$ cylinder. 
In the energy range $4 \le E \le 14$ meV, the measured cell transmission is 
well approximated by the superposition of three Gaussians,

\begin{eqnarray}
\label{transmission}
T(E)=&0.085+0.031\cdot \exp\left\{-\left(\frac{E-4.07}{0.283}\right)^2\right\} 
-0.035\cdot \exp\left\{-\left(\frac{E-5.3}{0.629}\right)^2\right\}\nonumber\\
 &+0.021\cdot \exp\left\{-\left(\frac{E-10.2}{3.58}\right)^2\right\} \;\; ,
\end{eqnarray}
as may be seen in Fig.~\ref{cellTrans}. Scattering intensities measured at 
$P=2.5$~GPa were subsequently corrected using the above expression for the 
energy-dependent transmission of the pressure cell normalized to  unity for
the constant-$\hbar\omega$ = 2 meV scan. Specifically, the actual count rate 
was divided by 
\begin{equation}
\tilde{T}(\hbar\omega)=
\frac{\sqrt{T(E_f+\hbar\omega)T(E_f)}}{\sqrt{T(6.24~{\rm meV})
T(4.24~{\rm meV})}}~
\approx~$10$\sqrt{T(E_f+\hbar\omega)T(E_f)} ,
\end{equation}
where $E_f$ is the final neutron energy, and $T(E)$ is given by 
Eq.~(\ref{transmission}). The actual pressure applied to the sample at low 
temperatures in our experiment was $P \approx$ 2.5~GPa, as determined from the 
measured change $\Delta a/a$=3.2$\times $10$^{-2}$ in the lattice 
constant of the NaCl sample \cite{cell}.

For the reference measurements at ambient pressure we extracted the microcell
with the sample from the pressure cell and placed it in a standard (ILL-type) 
flow cryostat. Its transmission for 5 meV neutrons measured the same way as 
above was 0.60(1), which can be attributed to incoherent scattering from the 
pressure-transmitting fluid and the NENP sample itself.

The  sample was mounted with its (0,1,0) and (1,0,3) reciprocal
lattice directions in the horizontal scattering plane. With this 
orientation, variations of the transverse (with respect to the spin chains) 
components of the wavevector transfer ${\bf Q}=(h,k,l )$ are coupled through
$h = l/3$. However, the  non-zero $h$ had negligible effect in our experiment. 
First, recalling that inelastic neutron scattering only probes spin fluctuations
polarized perpendicular to {\bf Q} \cite{Lovesey}, we note that the (1,0,0) 
direction was $\approx$ 80$^\circ$ out of the scattering plane. 
This means that at least 97$\%$ of the spin fluctuations polarized along 
{\bf a}$^*$ were always probed. Second, since the dispersion of the excitations 
along {\bf a}$^*$ has period 2 in $h$ \cite{NENP1}, it is quite negligible for 
$h < 0.2$ which is the case in present study. All constant-Q scans were performed 
at  $k = 1$, i.e. with wavevector transfer along the chain $\tilde{q}=\pi$.

\section{RESULTS AND DISCUSSION}

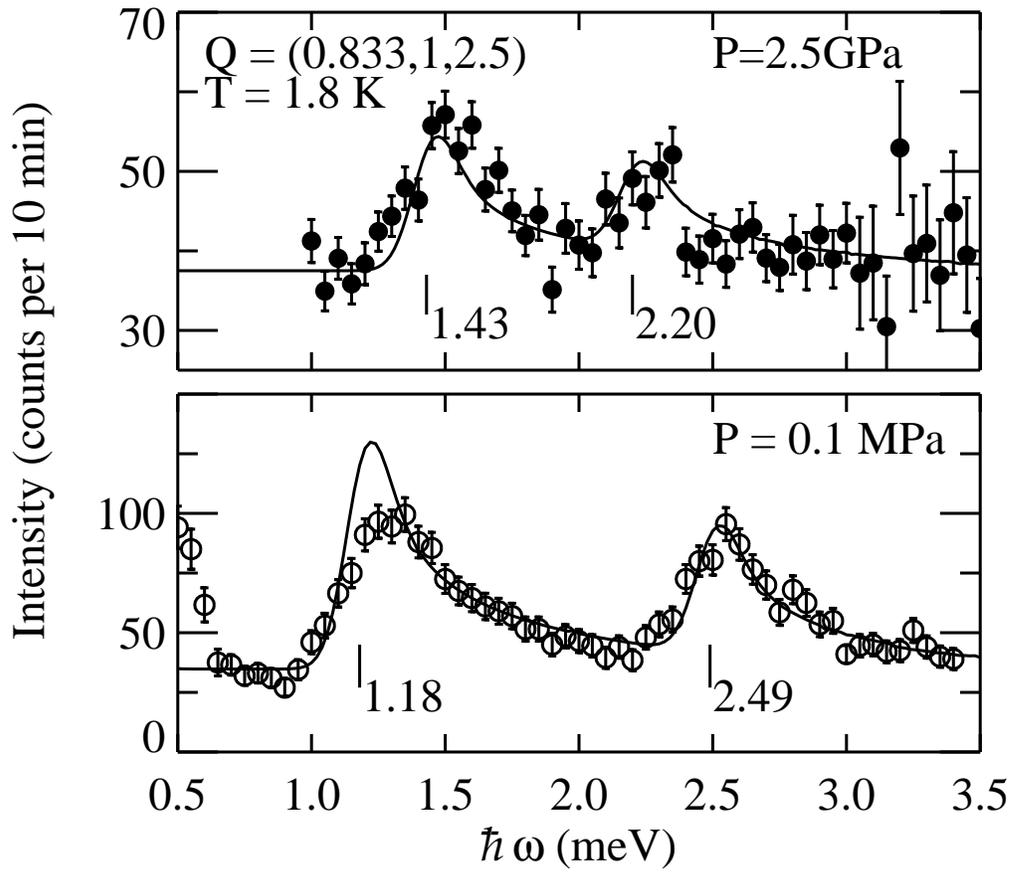
\begin{figure}
\caption{Constant-{\bf Q} scans at ${\bf Q}=(.833,1,2.5)$ picking up both 
transverse and longitudinal fluctuations at $P=2.5$~GPa (top) and $P=0.1$~MPa 
(bottom).}
\label{ConstQfig}
\end{figure}

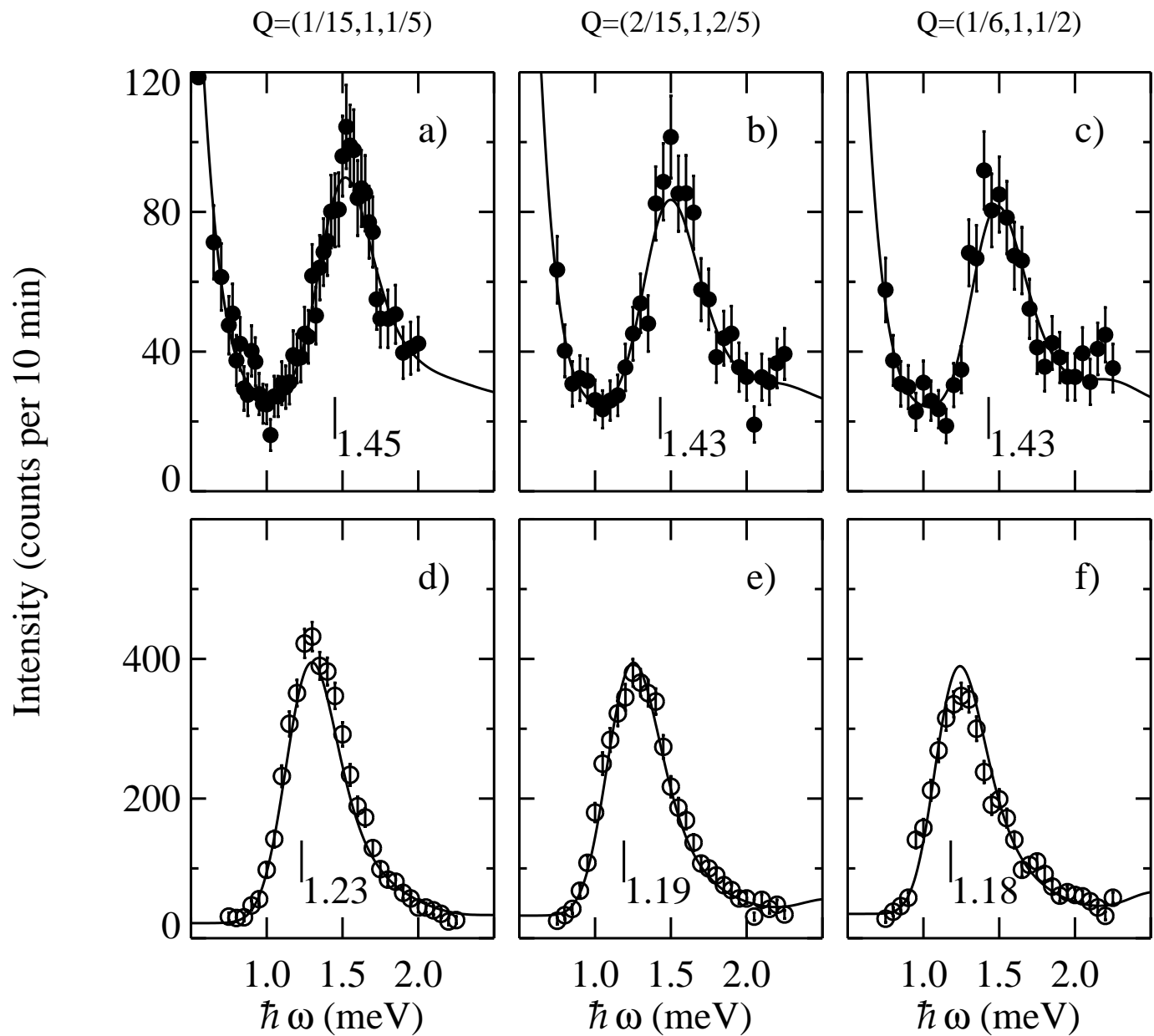
\begin{figure}
\caption{Scans at different $\tilde{q}_\perp$ probing the dispersion of
the transverse fluctuations with wavevector transfer perpendicular to the 
chains: (a)--(c) $P=2.5$~GPa; (d) and (e) $P=0.1$~MPa. }
\label{dispFig}
\end{figure}

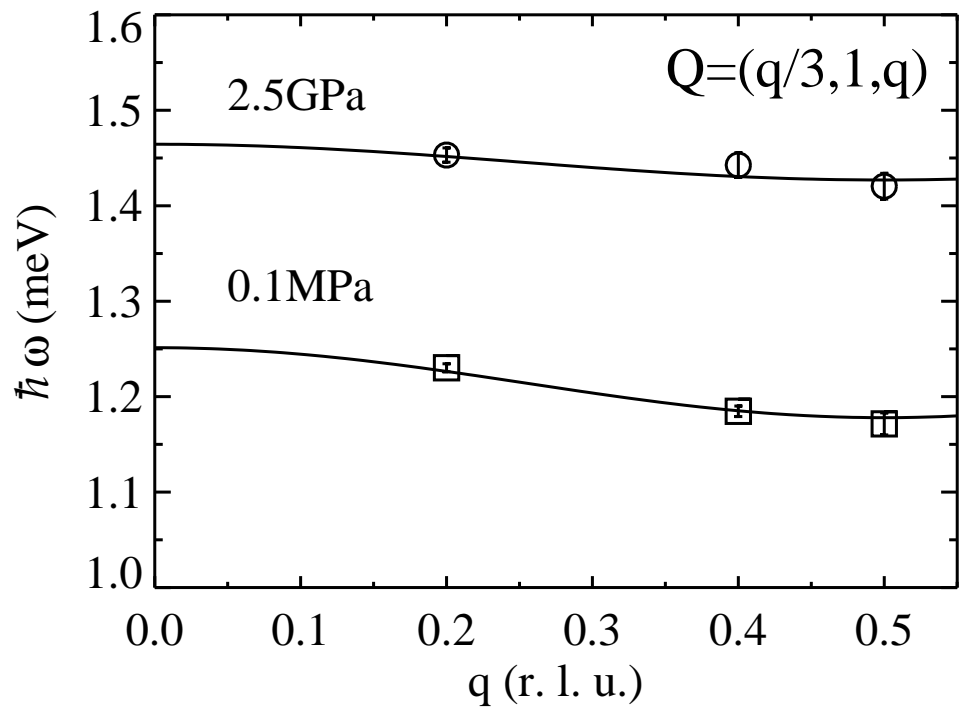
\begin{figure}
\caption{Dispersion curves for the transverse polarized gap mode obtained 
with Eq.~(\protect\ref{hw2}) using parameters from Table~1. Open symbols and 
error bars were obtained from independent unconstrained fits of each 
individual peak with the resolution convoluted lineshape. }
\label{c-disp}
\end{figure}

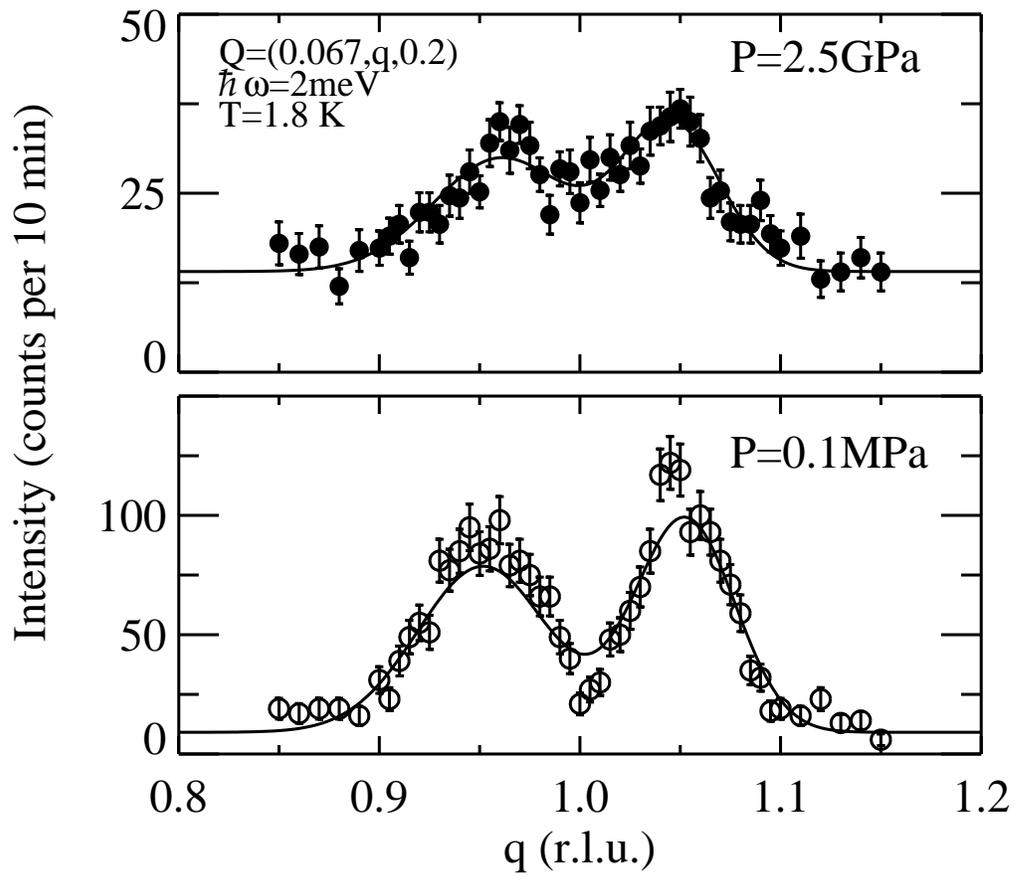
\begin{figure}
\caption{Constant-$\hbar\omega$ scans sensitive to the 
velocity of excitations 
along the chains at $P=2.5$~GPa (top) and $P=0.1$~MPa (bottom).}
\label{ConstEfig}
\end{figure}

Our experimental results are shown in Figs.~\ref{ConstQfig}--\ref{ConstEfig}. 
The scattering intensity is shown normalized  to 10 minutes counting time, 
although due to the low transmission of the pressure cell the bulk of the 
high-pressure data was counted three times this long to obtain adequate 
statistics. In Fig.~\ref{ConstQfig} we present energy scans at
{\bf Q}=(5/6,1,5/2). The large component of {\bf Q} transverse to the chains 
means that more than 90$\%$ of the intensity of the longitudinally polarized
(i.e. parallel to the chain axis) fluctuations is observed. At ambient pressure 
(0.1MPa) the corresponding gap is $\Delta_\|$= 2.49~meV, in agreement with 
previous studies \cite{NENP1,NENP2}. At $P = 2.5$~GPa this mode shifts to lower 
energy, and is found at $\Delta_\|$( 2.5 GPa) = 2.2~meV. The numbers quoted here 
are determined from the fits discussed below. In contrast, the energy of the 
transversely polarized mode, which at this {\bf Q} primarily contains 
fluctuations polarized along {\bf a}$^*$,  increases under pressure from 
$\Delta_\perp$ = 1.18~meV to $\Delta_\perp$(2.5GPa) = 1.43~meV. This increase 
of the energy gap in the magnetic excitations spectrum qualitatively agrees with 
the results of recent heat capacity measurements \cite{Ito} performed at pressures 
up to $\approx$ 0.5~GPa. 

Figure \ref{dispFig} shows scans that probe the dispersion of the transversely 
polarized modes with wavevector transfer perpendicular to the chains 
$q_\perp\approx 2 \pi l$. Presence of cell material in the beam cause much 
higher inelastic background and elastic scattering intensity in the 
high-pressure data. For $0\leq l \leq 0.5$, {\bf Q} makes angles between
0$^\circ$ and $33^\circ$ with the chain direction, so that between
100\% and 70$\%$ of the intensity of the {\bf a}$^*$-polarized transverse
fluctuations and up to 30$\%$ of the intensity of the longitudinal fluctuations 
is measured. The latter is seen as a small rise after the main peak in the
high-pressure data where the corresponding mode is centered at $\approx$ 2.2~meV.
As shown in Fig.~\ref{c-disp}, a small shift of $\Delta_\perp$ to higher 
energies as $l$ changes from 0.5 to 0 is clearly observed at ambient 
pressure. At P = 2.5~GPa, however, this shift is much smaller, and barely exceeds 
the experimental accuracy. Finally, one constant-$\hbar\omega$ = 2~meV
scan was performed at each pressure to probe the dispersion of the transverse
fluctuations with $\tilde{q}$ along the chains. These scans are shown in 
Fig.~\ref{ConstEfig}.

The solid curves shown in Figs.~\ref{ConstQfig}--\ref{ConstEfig}
were obtained as a result of a global fit to all the data at each pressure 
using the resolution-convoluted theoretical expression \cite{Lovesey} 
for the neutron scattering cross-section based on the single-mode
approximation (Eq.~(\ref{SMA})) for ${\cal S}^{\alpha\alpha}({\bf Q},\omega)$. 
To account for the dispersion perpendicular to the chains we used 
the dispersion relation

\begin{equation}
\label{hw2}
\hbar\omega_\alpha ({\bf Q}) = \sqrt{\Delta_\alpha^2 + v^2\sin^2
 \tilde{q}+A\sin^2\frac{\tilde{q}}{2} + (\Delta E_\perp)^2
\frac{1-\cos\tilde{q}\cos q_\perp}{2} } \;\; ,
\end{equation}
with $q_\perp = 2 \pi l$. This  was derived by adding in quadrature the energy
of the Haldane modes given by Eq.~(\ref{hw1}) with the transverse dispersion
obtained for magnons in spin-wave theory. We note that for $\tilde{q}$ close to 
$\pi$, Eq.~(\ref{hw2}) was obtained by Affleck \cite{Affleck} by
treating the array of weakly interacting Haldane chains on the basis of
the nonlinear sigma model. Except for the common parameters $\Delta_\|$, 
$\Delta_\perp$, $v$, $\Delta E_\perp$ in Eq.~(\ref{hw2}), for each scan only 
the individual constant background and gaussian intensity and width of the 
diffuse tail of incoherent scattering were varied in the global fit. Due 
to the limited range of  $\tilde{q}$ probed, the parameter A was fixed at
the value 34 meV$^2$ determined previously \cite{NENP2}. The results of
the fits for the parameters in the dispersion relation are summarized in 
Table~1. The values $\chi^2$=3.27 and $\chi^2$(2.5GPa)=0.95 once more indicate 
the validity of the single mode approximation (\ref{SMA}). 

It can easily be shown within the framework of perturbation theory \cite{NENP3} 
that the single-ion anisotropy term $D \sum_i(S_i^z)^2$ 
gives a splitting of the initially isotropic Haldane triplet  such 
that the average gap value remains unchanged: $\frac{1}{3}\Delta_\| + 
\frac{2}{3}\Delta_\perp = \Delta_H$. This fact can also be established  
with the symmetry arguments presented in the numerical study 
by Meshkov\cite{Meshkov}, where to  $\sim 1\%$ accuracy the  splitting was 
found to obey  

\begin{equation}
\label{splitting}
\Delta_\|=\Delta_H+\frac{4}{3}D\;\; ,\;\;\;\;\;\;
\Delta_\perp=\Delta_H-\frac{2}{3}D \;\; ,
\end{equation}
so that $\Delta_\|-\Delta_\perp=2D$. 
While we found that the average gap value $\Delta_H$ essentially did not 
change under pressure, the splitting of the gap decreased by more than 50$\%$. 
This implies that only $D$  is affected by hydrostatic pressure, while $J$ 
remains unchanged. Using Eq.~(\ref{splitting}) we arrive at the value 
$D$(2.5~GPa) = 0.09(7)$J$ for the anisotropy constant as compared to 
0.16(2)$J$ under normal conditions. This result can be given a simple 
interpretation: since the hydrostatic pressure is isotropic, 
it tends to pack charge distributions inside the crystal more 
symmetrically, thereby reducing the local anisotropy. From the transverse 
dispersion parameter $\Delta E_\perp\sim 4S\sqrt{2J'J}$, we estimate 
the exchange along the {\bf c}$^*$ direction to be $J'_c \approx 3\times 
10^{-4} J$ and $J'_c$(2.5GPa)$\approx 2\times 10^{-4} J $. This is even 
smaller than the value $J'_a \approx 8\times 10^{-4} J $ in the {\bf a}$^*$ 
direction measured previously \cite{NENP1}. Thus $J'_c$ appears to decrease 
with pressure, but this change is  at the limit of our experimental accuracy.

\section{CONCLUSIONS}

In conclusion, changes in the spin dynamics of the Haldane gap antiferromagnet 
Ni(C$_2$H$_8$N$_2$)$_2$NO$_2$ClO$_4$ (NENP) under  applied hydrostatic 
pressure of 2.5~GPa were studied by inelastic neutron scattering, and another 
point in the (D/J,J$'$/J) phase diagram of the S=1 spin system in the Haldane 
state has been attained experimentally. The principal effect of pressure was 
found to be a decrease by  a factor 1.7 of the single-ion anisotropy constant 
in the spin Hamiltonian of the Ni$^{2+}$ magnetic ions. This finding explains 
the increase with pressure of the effective spin gap in NENP observed in the 
recent heat capacity measurements \cite{Ito}, as being the result of the 
increasing energy of the lower doublet component of the Haldane triplet. 
The exchange constant in the $\hat{c}$-direction perpendicular to the chains 
was also measured for the first time.  It is smaller than the exchange along 
$\hat{a}$, and seems to decrease slightly with pressure. Thus the overall 
effect of applying hydrostatic pressure in NENP is not to move the system 
towards the phase transition to the N\'{e}el ordered ground state or the 
large-$D$ planar phase but to bring it closer to the isotropic 1D limit.

\section{ACKNOWLEDGEMENTS}

This work is based upon activities supported by National Science Foundation 
under agreement No. DMR-9413101. CB acknowledges support from the National 
Science Foundation through DMR-9453362. DHR acknowledges the support of the 
David and Lucile Packard Foundation.

\newpage

\begin{table}

\caption{ Summary of the dispersion parameters obtained from the global fits 
to data at two pressures; $\Delta_H$ is the average gap defined with 
Eq.~(\protect\ref{splitting}).}

\vspace{10mm}
\begin{tabular}{||c|c|c|c|c|c||}
%\hline
Pressure & $\Delta_\perp$ & $\Delta_\|$ & $\Delta_H$ & $v$ & $\Delta E_\perp $ \\
(Pa) & (meV) & (meV) & (meV) & (meV) & (meV)\\
\hline
10$^5$ & 1.180(4) & 2.49(1) & 1.62(1) & 9.34(20) & 0.43(3) \\
\hline
2.5$\cdot 10^9$ & 1.426(6) & 2.20(2) & 1.68(2) & 9.47(20) & 0.32(7) \\
%\hline
\end{tabular}
\vspace{3mm}

\end{table}

\end{document}